# Space charge dynamics in solid electrolytes with steric effect and Vegard stresses: resistive switching and ferroelectric-like hysteresis of electromechanical response

Anna N. Morozovska[1], Eugene A. Eliseev[2], Olexandr V. Varenik[3], Yunseok Kim[4], Evgheni Strelcov[4], Alexander Tselev[4], Nicholas V. Morozovsky[1], and Sergei V. Kalinin[4]

[1]Institute of Physics NAS of Ukraine, 46, pr. Nauki, Kiev, 03028 Ukraine

[2]Institute of Problems for Material Sciences, NAS of Ukraine, 3, Krjijanovskogo str., Kiev, 03028 Ukraine

[3] Taras Shevchenko Kyiv National University, Radiophysical Faculty
4, pr. Akademika Hlushkova, 03022 Kiev, Ukraine

[4]The Center for Nanophase Materials Sciences,
Oak Ridge National Laboratory, Oak Ridge, TN 37922

## Abstract

We performed self-consistent modelling of electrotransport and electromechanical response of solid electrolyte thin films allowing for steric effects of mobile charged defects (ions, protons or vacancies), electron degeneration and Vegard stresses. We establish correlations between the features of the space-charge dynamics, current-voltage and bending-voltage curves in the wide frequency range of applied electric voltage. The pronounced ferroelectric-like hysteresis of bending-voltage loops and current maxima on double hysteresis current-voltage loops appear for the electron-open electrodes. The double hysteresis loop with pronounced humps indicates the resistance switching of memristor-type. The switching occurs due to the strong coupling between electronic and ionic subsystem. The sharp meta-stable maximum of the electron density appears near one open electrode and moves to another one during the periodic change of applied voltage. Our results can explain the nature and correlation of electrical and mechanical memory effects in thin films of solid electrolytes. The analytical expression proving that the electrically induced bending of solid electrolyte films can be detected by interferometric methods is derived.

# 1. Introduction
## 1.1. Resistive switching: state of the art

Thin films of mixed ionic-electronic conductors (**MIEC**), such as solid electrolytes with mobile ions or vacancies, free electrons and/or holes, can display a reversible dynamics of the space charge layers that leads to a pronounced resistive switching between meta-stable states with high (**HR**) and low (**LR**) resistance [1, 2, 3, 4]. Electroresistive switching was observed in thin layers of semiconductor-ionics, solid electrolytes, doped semiconductors, e.g. $Ag_3AsS_3$ [5], Ti/PCMO/SRO and SRO/Nb:STO/Ag [1], $NiO_{1-x}$ [2], $Au/YBa_2Cu_3O_{7-x}/Au$ [6], Au nanocrystal-embedded $ZrO_2$ [7] and individual dislocations in $SrTiO_3$ [8]. Though the films are promising candidates for the nonvolatile resistive memory (**ReRAM**) devices, the physical principles of ReRAM operation, and in particular the role of the electromigration and diffusion of mobile ions and vacancies in the conductance switching, is far not clear.

Resistive switching observed in MIEC thin films typically belongs to memristor type [9, 10]. Dynamic current-voltage curves of a pronounced double loop shape is the distinctive feature of memristor (passive resistor with memory [11]). In general memristive systems cannot store energy, but they "remember" the total charge transfer due to the metastable changes of their conductance [12]. Modern theory of memristive switching originates from Strukov et al [13, 14, 15, 16], who demonstrated that memristive behaviour can be inherent to thin semiconductor films, when the drift-diffusion kinetic equations for electrons, holes and mobile donors/acceptors are strongly coupled, at that the memory resistance depends on the thickness ratio of the doped and pure regions of semiconductor.

Notably that the space charge dynamics in MIEC thin films were studied theoretically mostly in the framework of linear drift-diffusion Poisson-Planck-Nernst theory and diluted species approximation [13, 14, 17, 18]. However the understanding of the nonlinear processes and coupling mechanisms contribution to resistive switching and its theoretical description can be of great help to lift up the empirical methods of ReRAM design on a prognostic analytical level [1, 2, 13]. This research shows that the memristor-type switching between the HR and LR states can occur in electroded MIEC film due to the strong coupling between electronic and ionic subsystems mediated by the steric effects and chemical pressure.

## 1.2. Correlated electromechanical response of MIECs

One can expect a strong correlation between the electrophysical and electromechanical response of MIEC films [2]. Actually, the dynamic redistribution of mobile ions or vacancies concentration caused by electromigration (electric field-driven) and diffusion (concentration gradient-driven) mechanisms changes the lattice molar volume. The changes in the volume result in local electrochemical stresses, so called "Vegard stress" or "chemical pressure" [19, 20, 21]. The Vegard mechanism plays a decisive role in the origin and evolution of local strains caused by the point defect kinetics in solids [22, 23, 24, 25]. Note, that "steric" effect limits the maximal volume allowed per one ion, proton or vacancy [26, 27].

One- and two-dimensional analytical models of the linearized drift-diffusion kinetics in solid electrolytes with mixed electronic-ionic conductivity have been evolved [28, 29, 30, 31, 32]. The linear models [28-32], utilizing decoupling approximation for elastic problem solution and diluted electrolyte approximation for mobile charge species, proved that the Vegard stresses can give rise to the local mechano-electrochemical response in Li-containing MIECs and in charge defect surrounding of quantum paraelectrics. However the analytical results of the linear model, as decoupling approximation, are not applicable for MIEC film regions, where the electron accumulation and strong aggregation of mobile defects are pronounced. Such regions usually originate near the film interfaces under the increase of applied electric field more than the thermal activation field; corresponding voltage is typically not more than several tens of mV for ~100-nm films. Naturally, the expected correlation between the electrophysical and electromechanical response of MIEC films cannot be studied adequately within linear drift-diffusion theory, diluted electrolyte and decoupling approximations.

### 1.3. Research motivation and goals

Resuming available literature sources, the space charge dynamics in MIEC thin films were studied theoretically mostly in the framework of linear drift-diffusion theory and diluted electrolyte approximation [13-32]. Only a few theoretical models, which account for steric effects were proposed [26, 27, 33]. However the self-consistent calculations of the films electromechanical response allowing for the steric effects of mobile defects, electron/holes degeneration and Vegard stresses are absent to date. Though one can expect a pronounced memory effects and strong correlation between the current-voltage and strain-voltage response of the film, the latter was not studied theoretically. Intentions to fill the aforementioned gaps in the knowledge motivate us to formulate the following goals of our research.

- To perform self-consistent modelling of nonlinear electric transport and electromechanical response of solid electrolyte thin films allowing for steric effects of mobile defects (i.e. aggregation of ions, protons or vacancies), electron degeneration (i.e. using Fermi integral instead of Boltzmann statistics) and Vegard stresses.

- To calculate and analyze the space-charge dynamics, current-voltage and bending-voltage curves for different electric boundary conditions (electron-open, partially open or blocking, symmetric or asymmetric) and wide frequency range of applied electric voltage.
- To establish possible correlations between the features of space charge dynamics, current-voltage and bending-voltage curves. To prove that the bending can be detected experimentally.

## 2. Problem statement. Basic equations

Let us consider a heterostructure consisting of a thin top electrode, MIEC thin film (solid electrolyte) and bottom electrode [**Figure 1a**].

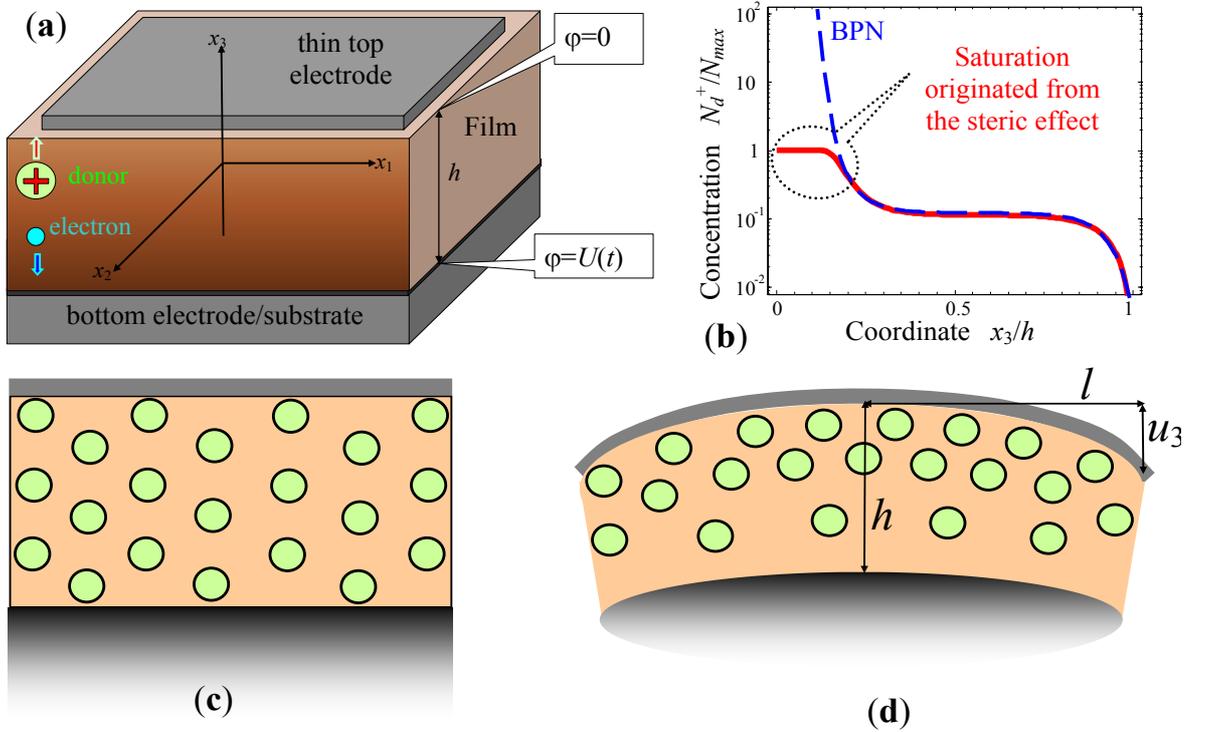

**Figure 1. (a)** Schematics of an electroded solid electrolyte film. Positively charged defects (cations or vacancies) and electrons are mobile in the film. **(b)** Typical difference in the spatial distribution of ionized defects calculated in the Boltzmann-Planck-Nernst (BPN) approximation (dashed curve) and allowing for the steric and Vegard effects (solid curve). **(c,d)** Appearance of the film bending. **(c)** Homogeneously distributed defects do not create any bending of the film. **(d)** Diffusion of ionized defects and their drift in electric field make the distribution inhomogeneous and corresponding chemical pressure leads to the film bending.

In the case all physical quantities depend only on the distance $x_3$ from the film-substrate interface (1D problem). Mobile positively charged point defects, oxygen vacancies or cations further regarded as *donors*, and free electrons are inherent to the film. The mobile charge carriers redistribution can create the internal electric field in the film, $E_3 = -\partial\varphi/\partial x_3$, and corresponding electric potential φ can be determined self-consistently from the Poisson equation:

$$\varepsilon_0\varepsilon\frac{\partial^2\varphi}{\partial x_3^2} = -e\left(Z_d N_d^+(\varphi) - n(\varphi)\right) \qquad (1)$$

Here $\varepsilon_0 = 8.85\times10^{-12}$ F/m is the dielectric permittivity of vacuum; ε is a dielectric permittivity of electrolyte [34], electron density is *n*, donor concentration is $N_d^+$, $e=1.6\times10^{-19}$ C is the electron charge, $Z_d$ is the donor charge. Let us regard that electric potential satisfy the fixed boundary conditions at the electrodes, $\varphi(0)=U(t)$ and $\varphi(h)=0$, which correspond to the electroded film of thickness *h*. The voltage *U* is applied to the top electrode.

Continuity equation for donor concentration $N_d^+$ is:

$$\frac{\partial N_d^+}{\partial t} + \frac{1}{eZ_d}\frac{\partial J_d}{\partial x_3} = 0, \qquad (2)$$

where the donor current $J_d$ is proportional to the gradients of the carrier electrochemical potentials levels $\zeta_d$ as $J_d = -eZ_d\eta_d N_d^+ (\partial\zeta_d/\partial x_3)$, where $\eta_d$ is the mobility coefficient that is regarded constant. The electrochemical potential level $\zeta_d$ is defined as [33]:

$$\zeta_d = -E_d - W_{ij}^d\sigma_{ij} + eZ_d\varphi + k_B T \ln\left(\frac{N_d^+}{N_d^0 - N_d^+}\right). \qquad (3)$$

Where $E_d$ is the donor level, elastic stress tensor is $\sigma_{ij}$, *T* is the absolute temperature, $k_B$ is a Boltzmann constant, $W_{ij}^d$ is the Vegard strain tensor (other name elastic dipole) [35, 36, 37]. Hereinafter the tensor is regarded diagonal $W_{ij}^d = W\delta_{ij}$ ($\delta_{ij}$ is delta Kroneker symbol). The absolute values of *W* for ABO$_3$ compounds can be estimated as $|W| \propto$ (1 – 50) Å$^3$ from the refs. [35-37]. The maximal possible concentration of donors ($N_d^0$) takes into account steric effects; namely $N_k^0 \equiv a^{-3}$, where $a^3$ is the maximal volume allowed per donor centre [26-27]. The steric effect limits the donor accumulation in the vicinity of film surfaces [**Figure 1b**]. Really, Equation (3) gives the donor concentration, $N_d^+ = N_d^0\left(1 - f\left(E_d + W_{ij}^d\sigma_{ij} - eZ_d\varphi + \zeta_d\right)\right)$; and thus $N_d^+ < N_d^0$ because the Fermi-Dirac distribution function $f(x) = (1+\exp(x/k_B T))^{-1}$ is always smaller than unity at finite temperatures. Note, that electrons are regarded size-less.

The electrodes are taken donor-blocking (i.e. impermeable for ions and vacancies), so the boundary conditions for the ionized donors current are $J_d(0) = J_d(h) = 0$.

Under negligibly small impact of the electron hopping process, continuity equation for electrons is:

$$\frac{\partial n}{\partial t} - \frac{1}{e}\frac{\partial J_3^e}{\partial x_3} = 0 \qquad (4)$$

Where the electron current $J_e = e\eta_e n(\partial \zeta_e / \partial x_3)$, $\eta_e$ is the electron mobility coefficient. Continuous approximation for the concentration of the electrons in the conduction band is consistent with the following expression for electro-chemical potential:

$$\zeta_e \approx E_C + k_B T F_{1/2}^{-1}\left(\frac{n(\varphi)}{N_C}\right) - e\varphi. \qquad (5)$$

The electro-chemical potential $\zeta_e$ tends to the Fermi energy level $E_F$ in equilibrium, $E_C$ is the bottom of conductive band, $F_{1/2}^{-1}$ is the function inverse to the Fermi integral $F_{1/2}(\xi) = \frac{2}{\sqrt{\pi}}\int_0^\infty \frac{\sqrt{\zeta}d\zeta}{1+\exp(\zeta - \xi)}$; effective density of states in the conductive band is $N_C = \left(\frac{m_n k_B T}{2\pi\hbar^2}\right)^{3/2}$, electron effective mass is $m_n$ [38]. Electron density can be calculated from Eq.(5) as $n = N_C F_{1/2}\left((e\varphi + \zeta_e - E_C)/k_B T\right)$. In Eq.(5) we neglect the deformation potential for the sake of simplicity, while it can be incorporated using the method [29, 31].

The boundary conditions for the electrons are taken in the linearized Chang-Jaffe (CJ) form, $J_e(0) = v_{e0}(n(0) - n_0)$ and $J_e(h) = -v_{eh}(n(h) - n_h)$, $v_e$ is a positive rate constant related with the surface recombination velocity [39]. CJ condition contains the continuous transition from the "open" electrode ($v_e \to \infty \Rightarrow n = n_0$) to the interface limited kinetics ($0 < v_e < \infty$) and "completely blocking" electrode ($v_e = 0$). Numerical values of $v_e$ are determined by the electrode and film material. If the rate constants are infinitely high, then the equilibrium electron concentrations at the contacts are pinned by the electrodes and independent on the applied voltage [40].

Using the dependencies for concentration of donors, $N_d^+ = N_d^0 f(-E_d - W\sigma + eZ_d\varphi - \zeta_d)$, and electrons, $n = N_C F_{1/2}\left((e\varphi + \zeta_e - E_C)/k_B T\right)$, on the electrochemical potentials $\zeta_{d,e}$, one can express the potentials as the functions of donor and electron chemical potentials [33]:

$$\mu_d = eZ_d\varphi - \zeta_d - W\sigma, \qquad \mu_e = e\varphi + \zeta_e, \qquad (6)$$

in the following way:

$$N_d^+ = N_d^0 f(\mu_d - E_d), \qquad n = N_C F_{1/2}((\mu_e - E_C)/k_B T). \qquad (7)$$

Then coupled Eqs. (2) and (4) become

$$\frac{\partial f(\mu_d - E_d)}{\partial t} - \frac{\partial}{\partial x_3}\left(\eta_d f(\mu_d - E_d)\frac{\partial(eZ_d\varphi - \mu_d - W\sigma)}{\partial x_3}\right) = 0 \qquad (8)$$

$$\frac{\partial}{\partial t} F_{1/2}\left(\frac{\mu_e - E_C}{k_B T}\right) - \frac{\partial}{\partial x_3}\left(\eta_e F_{1/2}\left(\frac{\mu_e - E_C}{k_B T}\right)\frac{\partial(\mu_e - e\varphi)}{\partial x_3}\right) = 0. \qquad (9)$$

In dimensionless variables, the equations for chemical potentials (8)-(9) coupled with Poisson equation for electric potential (1) and corresponding boundary conditions are listed below.

### 3. Vegard stresses cause the film bending

Homogeneously distributed defects do not create any bending of the film. Diffusion of ionized defects and their drift in electric field make the distribution inhomogeneous and corresponding chemical pressure leads to the film bending. The origin of the film bending caused by the local Vegard stresses is schematically shown in the **Figures 1c** and **1d**.

Equations of state, relating stress components, $\sigma_{ij}$, with strain components, $u_{ij}$, for film, containing inhomogeneous distribution of movable species with concentration $\delta N_d^+ = N_d^+ - \overline{N}_d^+$ are $u_{ij} = s_{ijkl}\sigma_{kl} + W_{ij}^d \delta N_d^+$. These equations should be supplemented by the mechanical equilibrium equations in the bulk and the normal stress components absence at the free surface of the system [41]. For the considered one-dimensional distributions, mechanical equilibrium equation is $\partial \sigma_{i3}/\partial x_3 = 0$ and the boundary condition $\sigma_{i3}|_S = 0$ ($i$=1, 2, 3) lead to the zero components $\sigma_{i3}$ throughout the sample. The conditions of elastic compatibility, $e_{ikl}e_{jmn}\partial u_{ln}/\partial x_{km} = 0$, are reduced to $\partial^2 u_{11}/\partial x_3^2 = 0$, $\partial^2 u_{22}/\partial x_3^2 = 0$. Meanwhile the distribution of $u_{33}$ can be arbitrary along $x_3$. Obvious solution is $u_{11} = u_{11}^{(0)} + \frac{x_3}{R_1}$ and $u_{22} = u_{22}^{(0)} + \frac{x_3}{R_2}$. Rigorously speaking, constants $u_{11}^{(0)}$, $u_{22}^{(0)}$, and curvature radii $R_1$ and $R_2$ should be found from the boundary conditions for zero normal stress on free surfaces and continuity of displacements on the interface between the film and substrate. To obtain analytical results, we used the Saint-Venant approximation for the boundary conditions on the film surfaces, $\int_0^h \sigma_{ii}(x_3)dx_3 = 0$ and $\int_0^h \sigma_{ii}(x_3)x_3 dx_3 = 0$ ($i$=1, 2), which means zero total force and total moment acting on the system in its plane. In this way approximate expressions for the film surface displacement components were derived [see **Appendix S1** in the **Suppl.Mat.**]:

$$u_1 = \left(\frac{W_{11}^d}{h}\int_0^h \delta N_d^+(z)dz + \frac{x_3}{R_1}\right)x_1, \quad u_2 = \left(\frac{W_{22}^d}{h}\int_0^h \delta N_d^+(z)dz + \frac{x_3}{R_2}\right)x_2, \quad (10a)$$

$$u_3 = \int_0^{x_3} u_{33}(\tilde{x}_3)d\tilde{x}_3 - \frac{1}{2}\left(\frac{x_1^2}{R_1} + \frac{x_2^2}{R_2}\right), \quad (10b)$$

$$\frac{1}{R_1} = \frac{12 W_{11}^d}{h^3}\int_0^h \left(z - \frac{h}{2}\right)\delta N_d^+(z)dz, \quad \frac{1}{R_2} = \frac{12 W_{22}^d}{h^3}\int_0^h \left(z - \frac{h}{2}\right)\delta N_d^+(z)dz. \quad (10c)$$

The last term in Eq.(10b) corresponds to the plate bending [**Figure 1d**]. Also one can see that $\frac{\partial^2 u_3}{\partial x_1^2} = \frac{1}{R_1}$, $\frac{\partial^2 u_3}{\partial x_2^2} = \frac{1}{R_2}$, so the values are "true" curvatures of the film surface. For the case $W_{11}^d = W_{22}^d = W$ the bending of the film can be calculated as the curvature of its surface,

$$\frac{1}{R} = \frac{1}{2}\left(\frac{\partial^2 u_3}{\partial x_1^2} + \frac{\partial^2 u_3}{\partial x_2^2}\right).$$

Notice, that the incident light beam deflection can be driven by the film swelling, and so by the value of $u_3 = l^2/R$, where $l$ is the film semi-width (see **Fig. 1d**). So it seems reasonable the detect experimentally the film bending stimulated by *ac* voltage using membrane-like film/substrate configuration operated in double-beam interferometer [42, 43] and X-ray diffraction [44, 45] systems.

Using Eqs.(10) and the relation $u_{ij} = s_{ijkl}\sigma_{kl} + W_{ij}^d \delta N_d^+$, the chemical stress included in Eqs.(6) is:

$$\sigma = \sigma_{11} + \sigma_{22} = -\frac{2W(N_d^+ - \bar{N}_d^+)}{(s_{11} + s_{12})}, \quad (11)$$

where the equilibrium concentration of ionized donors is $\bar{N}_d^+ = N_d^0(1 - f((E_d - E_F)/k_B T))$, $N_d^0$ is the maximal steric concentration of donors. Fermi level should be found from the condition $\mu_d = \mu_e = E_F$ substituted into the electroneutrality equation, $N_d^+ = n$, under the absence of electric potential ($\varphi = 0$) and elastic stress ($\sigma = 0$), or in its evident form, $N_C F_{1/2}((E_F - E_C)/k_B T) = N_d^0 f((E_F - E_d)/k_B T)$.

### 4. Results and their analyses

Numerical simulations of the space charge density, total space charge, potential, electric field, current-voltage (**I-V**) and film bending-voltage (**B-V**) curves were performed in MathLab for the coupled equations (1), (8)-(9) in dimensionless variables, as listed in **Appendix S2.** Boundary conditions correspond to the sinusoidal electric voltage applied to the donor-blocking electrodes. Symmetric or asymmetric, electron-blocking, partially or completely electron-open electrodes can be simulated by the rate constants $\xi_i$, which define the electron transparency of the donor-blocking electrodes. Quasi-static, slow or fast kinetic regimes are governed by dimensionless frequency $f$. These dimensionless parameters and others variables are listed in the **Table 1.**

Table 1. Dimensionless variables and parameters used in numerical simulations

| Quantity | Definition/designation | Numerical values (if applicable) |
|---|---|---|
| Dimensionless thickness | $\tilde{h} = h/L_D$ | $\tilde{h} = 20$, 40 and 10 |
| Debye screening length | $L_D = \sqrt{\dfrac{\varepsilon_0 \varepsilon_{33}^b k_B T}{e^2 \overline{N}_d^+}}$ | ~ nm, depends on $\overline{N}_d^+$ in a self-consistent manner |
| Dimensionless time | $\tilde{t} = t/t_e$ | variable |
| Dimensionless frequency of applied voltage | $f = t_e \omega / 2\pi$ | $10^{-1}$, $10^{-2}$ (very high and high frequencies denoted as $f_1$ and $f_2$)<br>$10^{-3}$, $3\times 10^{-4}$, $10^{-4}$ (intermediate frequencies denoted as $f_3 - f_5$)<br>$10^{-5}$, $10^{-6}$ (low and very low frequencies denoted as $f_6$ and $f_7$) |
| Characteristic electronic and donor times | $t_e = L_D^2 / (e \eta_e k_B T)$,<br>$t_d = L_D^2 / (e \eta_d k_B T)$ | $t_d / t_e \equiv \eta_e / \eta_d = 10, 10^2, 10^3$ |
| Dimensionless electric field | $\tilde{E} = e L_D E / (h k_B T)$ | variable |
| Dimensionless donor concentration | $\tilde{N} = N_d^+ / \overline{N}_d^+$ | variable |
| Dimensionless electron density | $\tilde{n} = n / \overline{N}_d^+$ | variable |
| Dimensionless total current | $\tilde{J} = \dfrac{L_D J}{e \overline{N}_d^+ \eta_e k_B T}$,<br>$J = J_e + J_d + (\partial E_3 / \partial t)$ | variable<br>The total current is coordinate-independent as anticipated |
| Dimensionless rate constant | $\tilde{\xi}_{0,h} = \dfrac{L_D v_{e0,h}}{e \eta_e k_B T}$ | $>10^3$ (electron-open electrodes)<br>1 (partially electron-open electrodes)<br>$10^{-3}$ (weakly electron-open electrodes)<br>0 (electron-blocking electrode) |
| Dimensionless bending | $\dfrac{1}{W \overline{N}_d^+} \dfrac{h}{R}$ | variable |

Space charge distribution, I-V and B-V curves were calculated for the donor-blocking and electron-blocking, partially or completely electron-open electrodes, high, intermediate and low frequencies of applied voltage. The bending effect is defined by the curvature of the film surface. Typical results are presented in the **Figures 2-7**. Rather versatile features of the space charge dynamics, I-V and B-V curves are summarized in the **subsections 4.2** and **4.3** respectively.

### 4.2. Typical features of the space charge dynamics

In the case of donor-blocking and electron-blocking electrodes [**Figure 2**]**,** the transfer of ionized donors and electrons in the opposite directions leads to their accumulation in under-electrode film regions (in accordance with carrier sign and polarity of applied ac voltage). This corresponds to the periodical successive nucleation, growth and annihilation of *dynamic ionic-electron quasi-dipole* and so to the formation and reorientation of corresponding polarization. This allows us to compare obtained characteristics with the ones typical for ferroelectrics. Steric effects leading to the plateau on the ionized donor distribution in under-electrode film regions are pronounced. As it should be, the total amount of electrons and donors remains constant within the film at all voltages due to blocking electrodes.

If one donor-blocking electrode is electron-open and another one is electron-blocking [**Figure 3**], the strong accumulation of ionized donors or electrons occurs in the film region adjacent to the electron-blocking electrode depending on the polarity of applied voltage. Rather weak accumulation of the space charges occurs in the film region adjacent to the electron-open electrode during the period of applied ac voltage. The total amount of donors remains constant in the film. Total amount of electrons as the function of applied voltage amplitude has strongly asymmetric step-like shape at low frequencies and transforms into irregular leaf-like loops under the frequency increase; then it becomes a flattened elliptic loop at high frequency [**Figure 6b**]. The periodic spatial variation of electron amount (below called modulation depth) is very high at low frequencies (up to $10^4$ times), but decreases strongly under the frequency increase. At high frequencies it oscillates in the vicinity of unity. The film is enriched by electrons at positive voltages and depleted by them at negative ones (electrode polarity effect).

In the case of donor-blocking and electron-open electrodes [**Figure 4**] each near-electrode film region accumulates both ionized donors and electrons, but the values and profiles of corresponding space charges are different. Exactly the profile difference determines the total charge and electric field distribution across the film. A pronounced maximum on the electron density appeared near one electrode and moves to the other one under the change of applied voltage polarity. Total amount of electrons as the function of applied voltage amplitude has symmetric V-like shape at low frequencies and transforms into a butterfly-like hysteresis loop under the frequency increase; and then acquires a two-leaved shape at high frequency [**Figure 7b**]. The electron amount modulation depth is about 30% at low frequencies and rapidly decreases to 1-3% under the frequency increase. At high frequencies it oscillates in the immediate vicinity of unity. The total amount of donors remains constant.

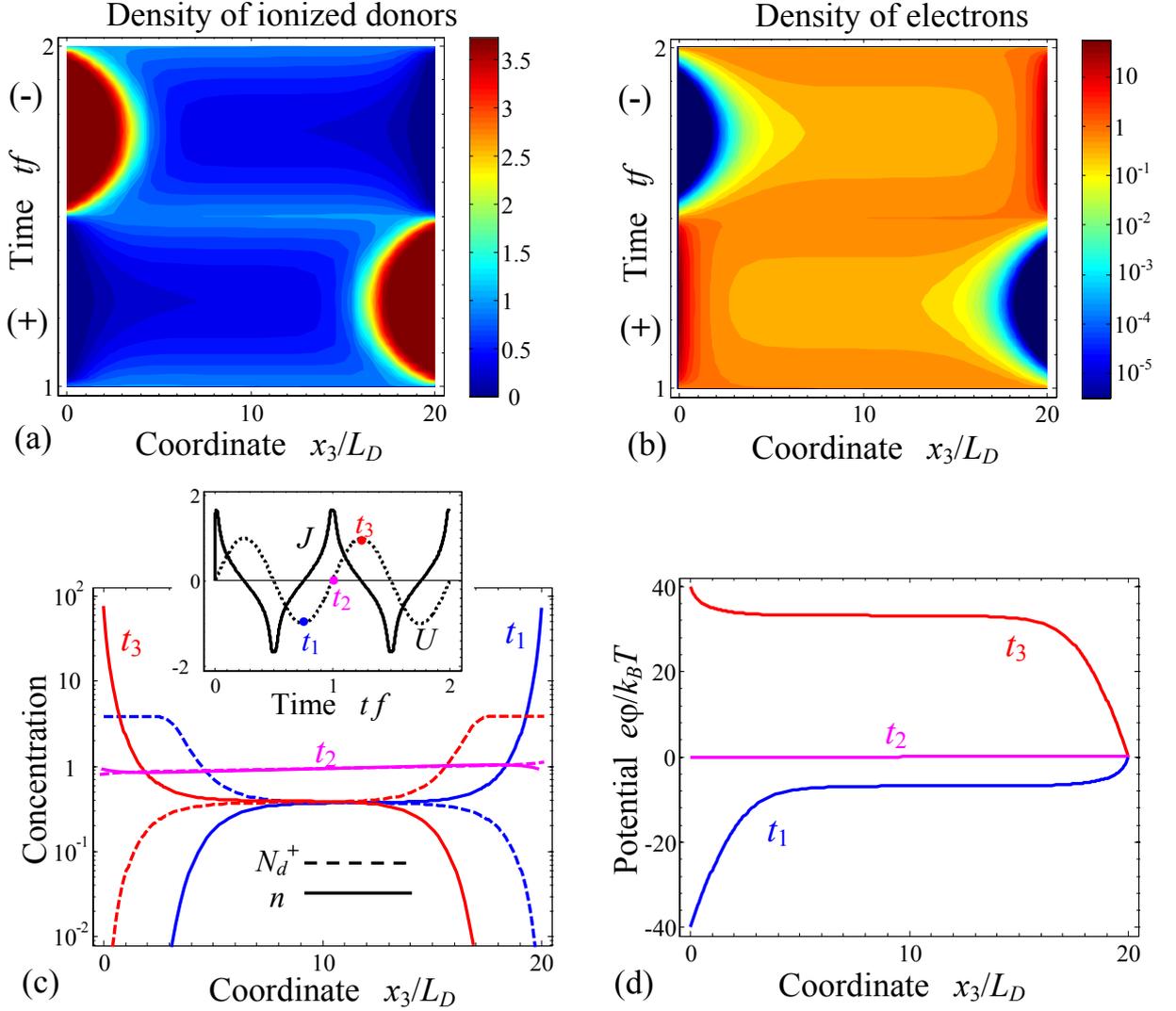

**Figure 2. Space charge dynamic profiles calculated for donor-blocking and electron-blocking electrodes.** Changes of the ionized donors $N_d^+/\overline{N}_d^+$ **(a)** and electrons $n/\overline{N}_d^+$ **(b)** distribution during one period of applied ac voltage. Electron $n/\overline{N}_d^+$, donor $N_d^+/\overline{N}_d^+$ **(c)** and potential $e\varphi/k_BT$ **(d)** profiles calculated for three successive moments of time $t_1$, $t_2$ and $t_3$ indicated in the inset to plot (c). Dimensionless frequency $f = 10^{-6}$, applied voltage amplitude $eU/k_BT = 40$, film thickness $\tilde{h} = 20$, ratio $t_d/t_e = 100$, $E_C = 0.0$ eV, $E_d = -0.1$ eV, donor concentration $N_d^0 = 10^{24}$ m$^{-3}$, Vegard coefficient $W = 10$ Å$^3$, T=298 K, permittivity $\varepsilon = 5$.

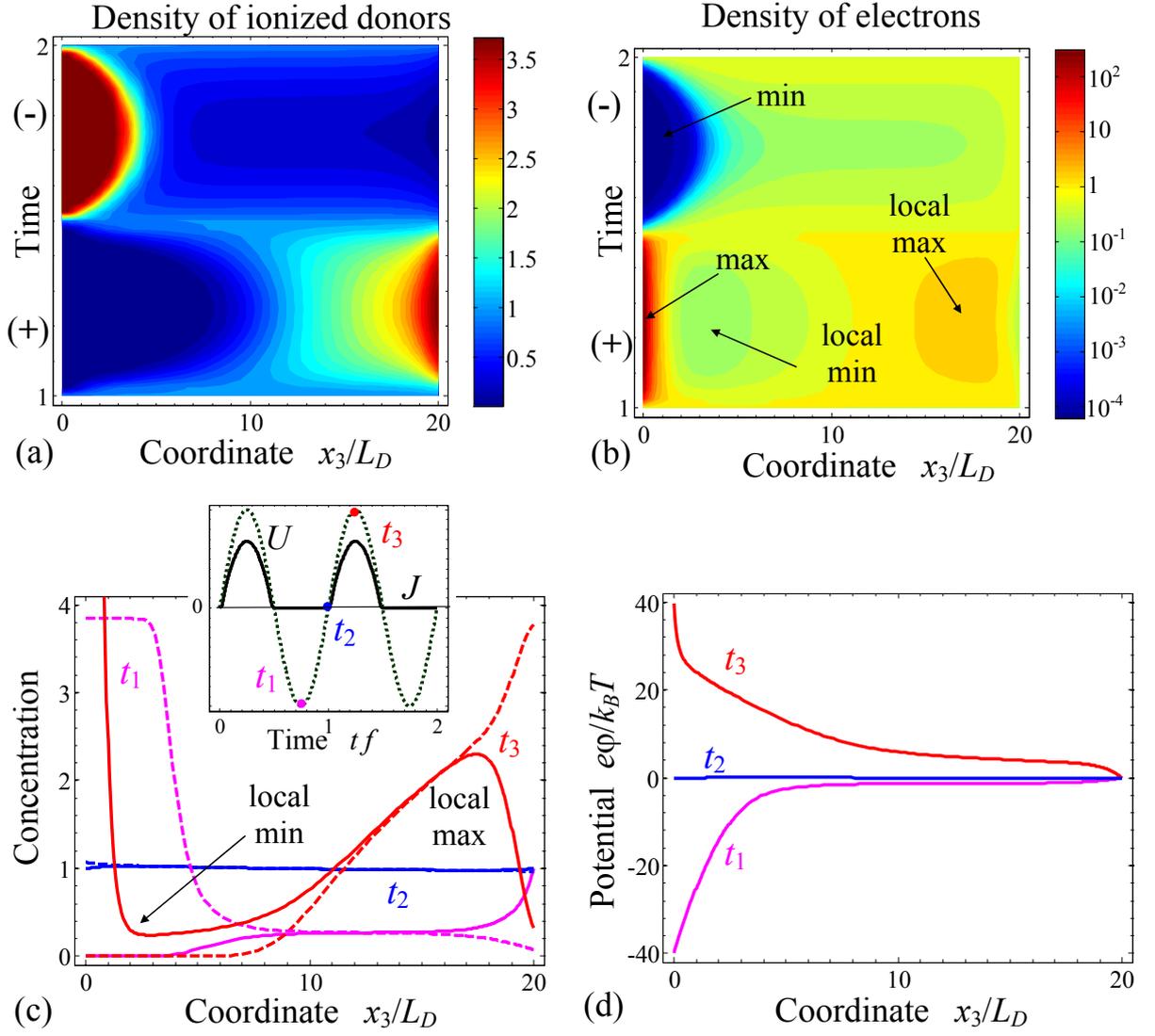

**Figure 3. Space charge dynamic profiles calculated for asymmetric donor-blocking electrodes. One electrode is almost electron-blocking ($\widetilde{\xi}_0=10^{-3}$), another is partially electron open ($\widetilde{\xi}_h=1$).** Changes of the ionized donors $N_d^+/\overline{N}_d^+$ (a) and electrons $n/\overline{N}_d^+$ (b) distribution during one period of applied ac voltage. Electron $n/\overline{N}_d^+$, donor $N_d^+/\overline{N}_d^+$ (c) and potential $e\varphi/k_BT$ (d) profiles calculated for three successive moments of time $t_1$, $t_2$ and $t_3$ indicated in the inset to plot (c). Other parameters are the same as in the figure 2.

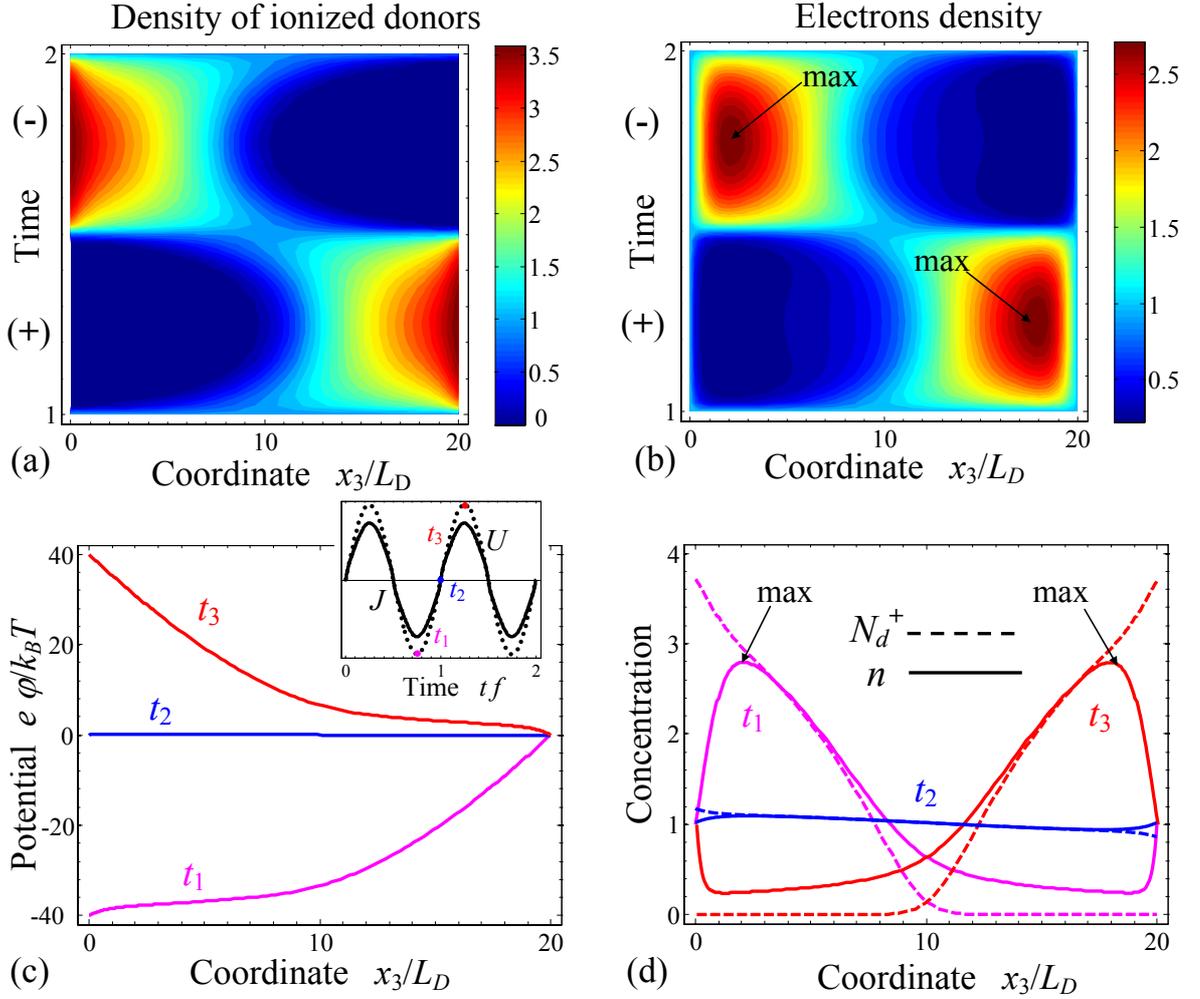

**Figure 4. Space charge dynamic profiles calculated for symmetric donor-blocking and electron-open electrodes.** Changes of the ionized donor $N_d^+/\overline{N}_d^+$ **(a)** and electron $n/\overline{N}_d^+$ **(b)** distribution during one period of applied ac voltage. Potential $e\varphi/k_B T$ **(c)**, electron $n/\overline{N}_d^+$ and donor $N_d^+/\overline{N}_d^+$ **(d)** profiles calculated for three successive moments of time $t_1$, $t_2$ and $t_3$ indicated in the inset to plot (c). Other parameters are the same as in the figure 2.

### 4.3. Typical features of I-V and B-V curves

In the case of both donor- and electron-blocking electrodes and the low enough frequency, I-V loop is rhomb-like and has slightly asymmetric single diffuse maximum around zero voltage [**Figure 5a**]. At very low frequencies ($f_7$ and $f_6$) the current amplitude decreases linearly under the frequency decrease. The frequency increase leads to considerable increase of the current amplitude. Inclined diamond-like I-V loops with rounded corners and pronounced asymmetric maxima with voltage position slightly shifted relatively to zero voltage are typical for intermediate frequencies $f_5 - f_3$. The maxima value and voltage shift increases under the frequency increase from $f_5$ to $f_3$. With further increase of the frequency the maxima are blurred and I-V loop transforms into a bone-like loop with two diffuse maxima at nonzero voltages. Then swelling up and transformation in a tilted elliptic loop occurs under the frequency increase from $f_2$ to $f_1$.

If the electrodes are donor- and electron-blocking and the frequency is low, B-V loop is symmetric and has a paraelectric-like hyperbolic tangent shape [**Figure 5b**]. The film bending saturates under the ac voltage amplitude increase. Under the frequency increase in the intermediate frequencies range the appreciable increase of the loop width and noticeable decrease of loop height take place that corresponds to the transformation of relaxor ferroelectric-like to ferroelectric-like hysteresis loop. Under subsequent frequency increase the transformation to elliptic loop is accompanied by considerable decrease of the loop height.

If both electrodes are donor-blocking, but one of them is electron-open and another one is electron-blocking [**Figure 6c**] and the frequency is low, I-V loop is strongly asymmetric with very small current on the negative branch and saturation region on the positive branch. Strongly asymmetric I-V loops with hysteresis-like positive voltage branch and saw-tooth-like current maximum on the negative branch are typical for intermediate frequencies $f_5 - f_3$. The maximum value increases and then decreases under the frequency increase. Swelling up of asymmetric loop and its transformation in a tilted elliptic-like loop occurs under the frequency increase from $f_2$ to $f_1$.

If one of the donor-blocking electrodes is electron-open, and another one is electron-blocking and the frequency is low B-V loop is noticeably asymmetric with hyperbolic tangent shape of negative high bending branch and saturated positive low bending branch [**Figure 6d**]. For intermediate frequencies the transformation of asymmetric double hysteretic B-V-loop into the tilted drop-like one takes place. At that the loop opening increases under the frequency increase . Under subsequent frequency increase the transformation of asymmetric loop to elliptic one is accompanied by the height decrease.

In the case of donor-blocking and electron-open electrodes [**Figure 7c**] and low frequencies, I-V loop branches are quasi-linear at low voltages and weakly superlinear at high voltages. For the frequencies $f_7$ and $f_6$ two current maxima are noticeable at symmetric small voltage positions. For intermediate frequencies $f_5 - f_3$ the current maxima become more pronounced and shift to higher voltages under the frequency increase. Bow-like double-hysteresis regions correspond to the transition between HR and LH states. With the frequency increase the I-V loop swells up, loses the maxima and becomes a tilted ellipse.

If both electrodes are donor-blocking and electron-open [**Figure 7d**] and the frequency is low enough, symmetric B-V loop has a paraelectric-like shape. The bending value saturates under the voltage increase. Transformation of the paraelectric-like loop into a symmetric ferroelectric-like hysteretic loop occurs under the frequency increase. Here we used the term "ferroelectric-like", because the loop shape resembles hysteretic loop of polarization reversal in ferroelectrics. Transformation of the ferroelectric-like loop, which width increases and height decreases under the frequency increase, to elliptic loop occurs under further increase of the frequency. At high frequencies the elliptic loop height appreciably decreases under the frequency increase.

To resume the subsection, the most striking features we see for the most common case of electron-open electrodes are the pronounced current maxima on the I-V curves at low and intermediate frequencies [**Figure 7c**] and saturated hysteresis of B-V curves [**Figures 5b** and **7d**], which resemble the polarization reversal in ferroelectrics. The current maxima voltage positions on I-V curves are very close to zero bending voltages on B-V hysteresis loop [compare **Figures 7c** and **7d**]. Zero bending voltages was defined similarly to the coercive voltage of ferroelectrics [46]. In the case of very high frequency the I-V and B-V loops have elliptic form, and their magnitude is weakly dependent on the boundary conditions.

Pronounced double hysteresis of I-V loop indicates the possibility of resistive switching between HR and LR states, i.e. the memristive effect. Our modelling reveals that the memristor-type switching between the HR and LR states occurs due to the strong coupling between electronic and ionic charge transfer appeared under the changes of voltage applied to the film electrodes. The meta-stable pronounced maximum of the electron density forms near one of the electrodes. Then it moves from the one electrode to another during the periodic change of applied *ac* voltage [**Figure 4b** and **4d**]. The metastability leads to the butterfly-shaped loops of the total electron amount [**Figure 7c**] and to the double hysteresis of the dynamic I-V loop [**Figure 7d**]. However we cannot speak about the pronounced memristor effect because the difference between the LR and HR states, defined as the slope change from the I-V loop, is not very high (estimation from Fig.8b gives HR/LR ratio ~10). Despite the warning the double hysteresis loop shape resembles the results of Refs.[1, 13, 14].

Notably, a pronounced hysteresis of the surface displacement amplitude as a function of pulse voltage is typically observed in MIECs by Electrochemical Strain Microscopy (ESM) [47, 48, 49, 50, 51, 52, 53], when the voltage sweep frequency is close to the inverse characteristic diffusion time in the probed volume [47-53]. At that nucleation voltages (corresponding to the inflection points on the forward and reverse curves [54]) do not depend on the bias applied to the tip, strongly suggesting the rate-independent thermodynamically limited nucleation-like phenomena. Available theoretical models of ESM response are linear [28-32]. The linear models proved that the coupling between ionic redistribution and Vegard strains give rise to the ESM response in these materials and describe adequately it frequency spectrum, but the observed ferroelectric-like shape of hysteresis loops remained mainly unexplained. One-dimensional [28, 29] and two-dimensional analytical models [30] of linearized diffusion kinetics in ESM, and 2D analytical model that considers linearized drift-diffusion kinetics [31, 32] have been evolved and result to the elliptic loop shape with "coercive" voltage determined by applied bias. In contrast to the previous works [28-32], this research describes the ferroelectric-like saturated shape of hysteresis loops; it is the first self-consistent modelling of the local mechano-electro-chemical response of solid electrolytes kinetics allowing for the strong nonlinearity and steric effect inherent to the system.

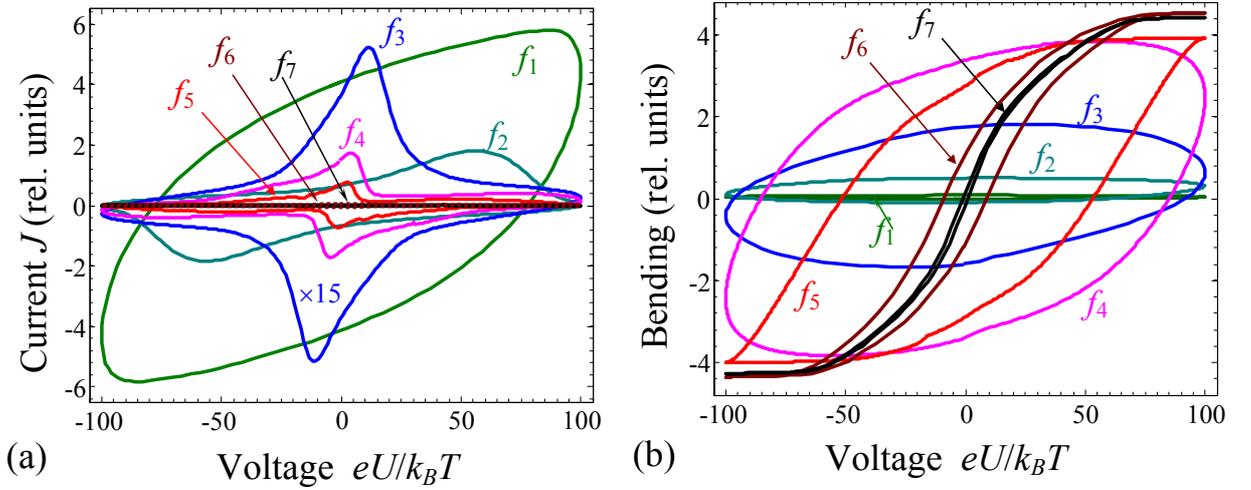

**Figure 5. I-V- (a) and B-V- (b) curves calculated for donor-blocking and electron-blocking electrodes.** I-V curves amplitude calculated for the frequencies $f_3 - f_7$ are multiplied by the factor 15. Dimensionless frequencies of applied voltage are $f = 10^{-1}$, $10^{-2}$, $10^{-3}$, $3\times10^{-4}$, $10^{-4}$, $10^{-5}$, $10^{-6}$ (curves $f_1 - f_7$ respectively). Other parameters are the same as in the figure 2.

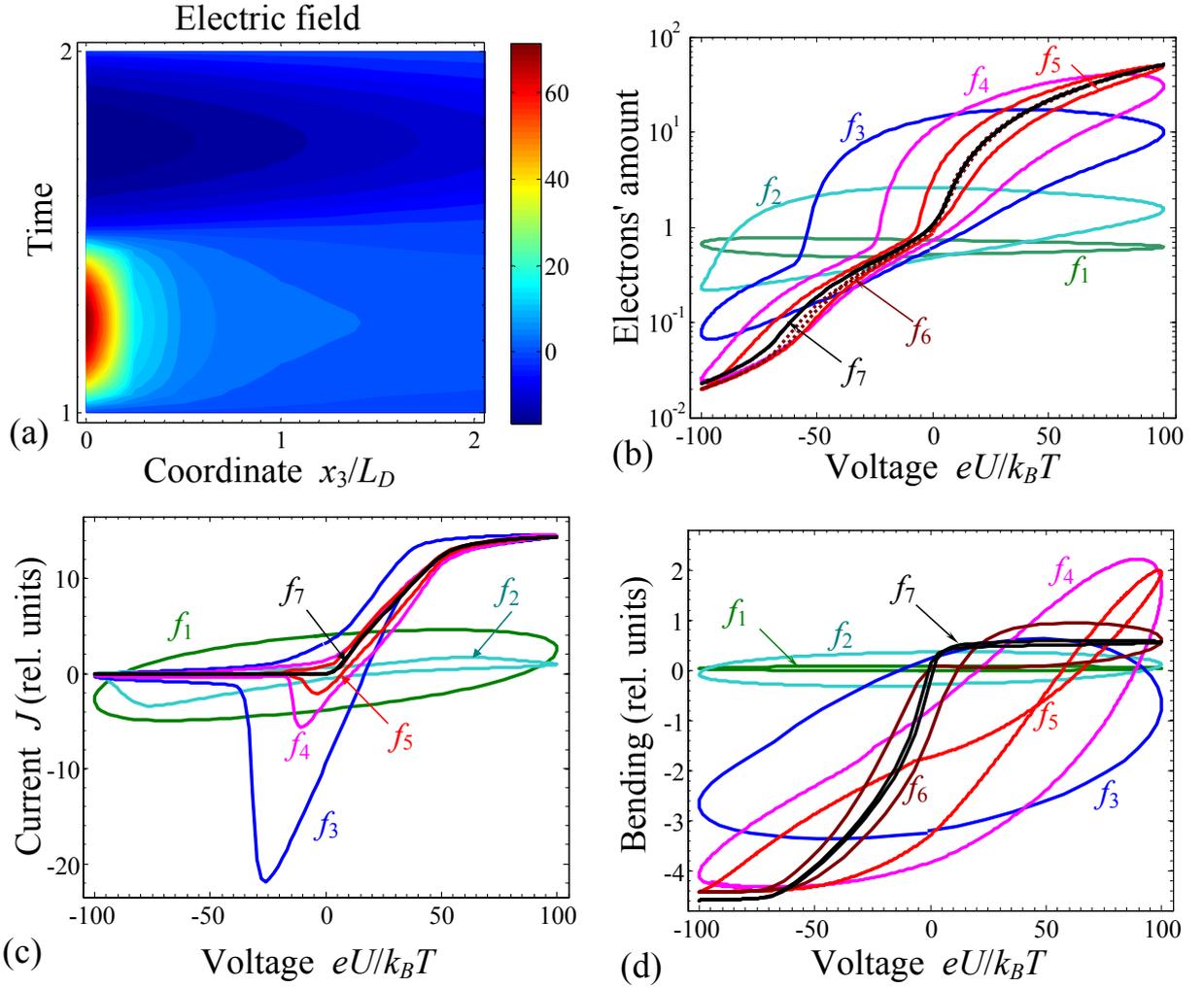

**Figure 6.** Electro-physical and electro-mechanical characteristics calculated for asymmetric donor-blocking electrodes: one electrode is almost electron-blocking ($\tilde{\xi}_0 = 10^{-3}$), another is partially electron open ($\tilde{\xi}_h = 1$). **(a)** Changes of the electric field distribution calculated during one period of applied ac voltage for the voltage amplitude $eU/k_BT = 40$ and lowest frequency $f = 10^{-6}$. **(b)** Total number of electrons as the function of applied voltage amplitude. **(c)** I-V and **(d)** B-V curves. Other parameters are the same as in the figures 2.

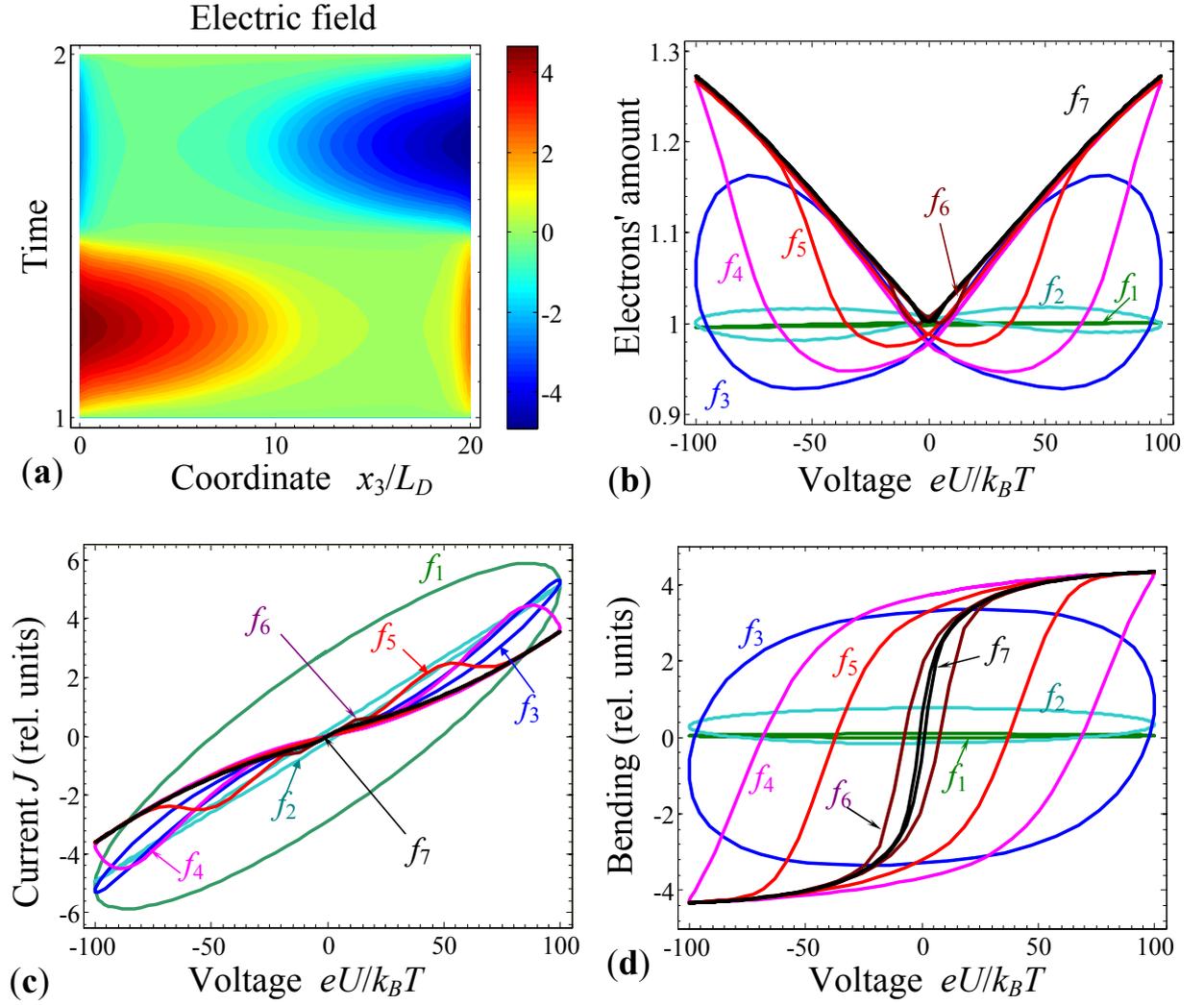

**Figure 7. Electro-physical and electro-mechanical characteristics calculated for donor-blocking and electron-open electrodes. (a)** Changes of the electric field distribution calculated during one period of applied ac voltage for the voltage amplitude $eU/k_BT = 40$ and lowest frequency $f = 10^{-6}$. **(b)** Total number of electrons as the function of applied voltage amplitude. **(c)** I-V and **(d)** B-V curves. Dimensionless frequencies of applied voltage are $f = 10^{-1}$, $10^{-2}$, $10^{-3}$, $3 \times 10^{-4}$, $10^{-4}$, $10^{-5}$, $10^{-6}$ (curves $f_1 - f_7$ respectively). Other parameters are the same as in the figure 2.

### 4.4. Dynamic nature of the memristor-type I-V and ferroelectric-like B-V curves

**Figures 8** demonstrate dynamic nature of revealed memristor-type (for I-V) and ferroelectric-like (for B-V) behaviour. The maxima on I-V curves and ferroelectric-like hysteresis shape of B-V curves originate from the strong retarding of the ionic subsystem with respect to the electron one. Namely the current maxima position, height and width is determined by the ratio of electronic to ionic relaxation times, $t_d/t_e$. At high ratio, $t_d/t_e = 10^3$, the maxima are pronounced at low frequency and for the thicker film. Zero-bending voltage of the B-V hysteresis strongly increases with $t_d/t_e$ ratio increase. For all considered $t_d/t_e$ ratios and different film thickness $\tilde{h}$ the current maxima position on I-V loop is very close to the zero-bending voltage of the B-V loop.

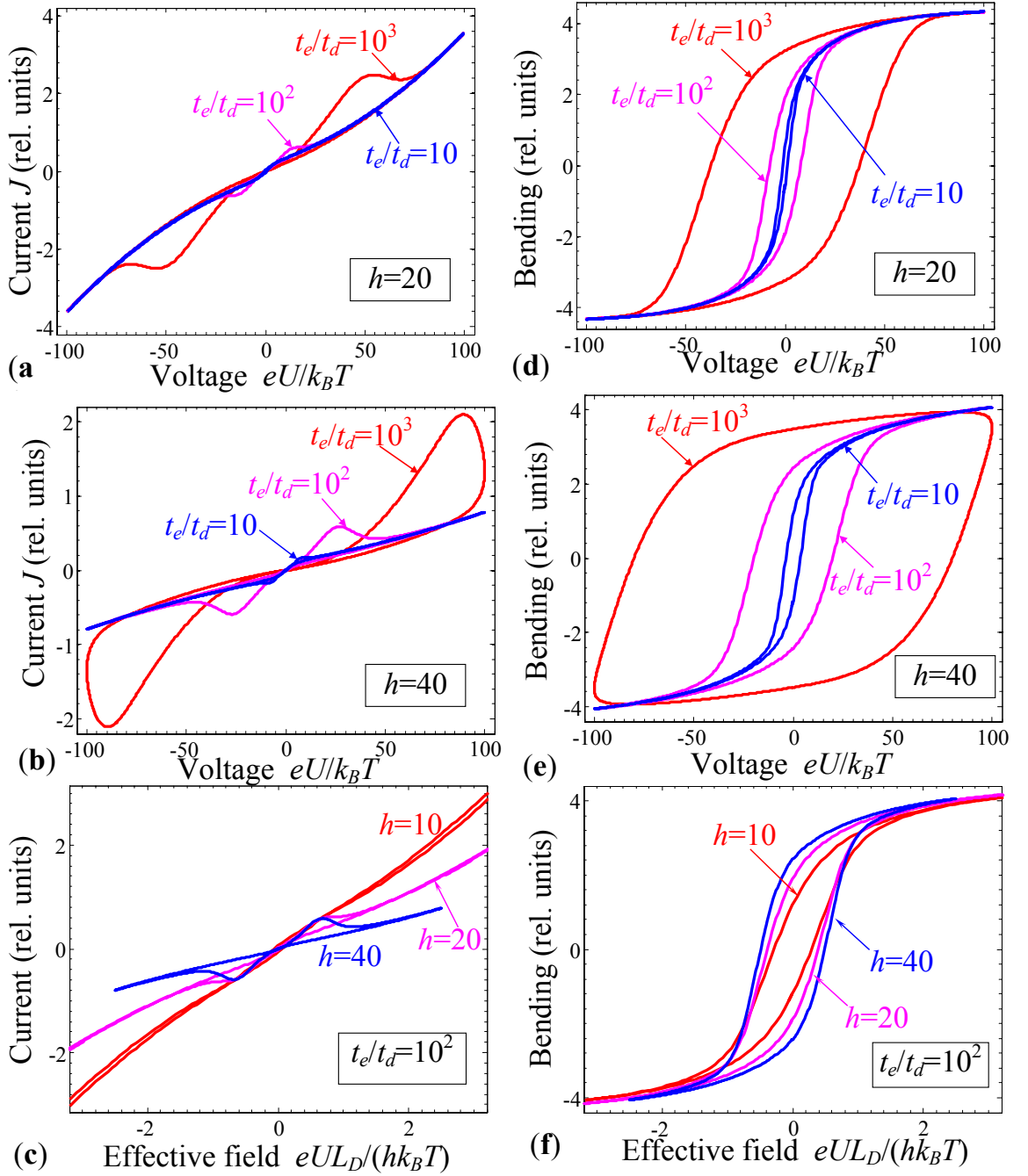

**Figure 8. I-V (a, b, c) and B-V (d, e, f) curves calculated for donor-blocking and electron-open electrodes.** Dimensionless frequencies of applied voltage $f = 10^{-5}$, film thickness $\tilde{h} = 20$ (a, d) and $\tilde{h} = 40$ (b, e), $\tilde{h} = 10, 20, 40$ (c, f), ratio $t_d/t_e = 10$, $10^2$, $10^3$ (labels near the curves). Other parameters are the same as in the figure 2.

Analyses of the obtained results establish the following correlation between the I-V and B-V curves. Independently on the donor-blocking electrodes properties the curves acquire elliptic shape under the frequency increase, but the B-V curves become elliptic at essentially lower frequencies than the I-V curves. The B-V curves behavior resembles the total electric charge hysteresis appeared during the spontaneous polarization reversal in ferroelectric materials. In particular, for the completely blocking electrodes the I-V-loops look like those of capacitor $BaTiO_3$-based ceramics in shallow paraelectric phase [55]. When the donor-blocking electrodes are also electron-blocking the voltage position of the current maxima corresponds to the zero-bending voltage on the B-V loops at low and intermediate frequencies. At very high frequency the ionic contributions to the bending and current vanish. When one electrode is electron-open, another is electron-blocking the I-V and B-V curves demonstrate a pronounced asymmetry at low and intermediate frequencies. Bending and current loops are open at negative voltages for intermediate frequencies. For donor-blocking and electron-open electrodes the electric current maxima are typical. Their position is very close to the zero bending voltages on the B-V hysteresis loop. Pronounced current maxima on the double hysteresis I-V loop indicate the possibility of resistive switching between low and high conductive states, i.e. the memristor-type effect.

**5. Summary**

Self-consistent modelling of electrical transport and electro-mechanical response of solid electrolyte thin films was performed allowing for steric effects of mobile ions or vacancies, electron degeneration and Vegard coupling. Correlations between the features of the space-charge dynamics, current-voltage and bending-voltage curves were established for different electric boundary conditions in the wide frequency range of applied electric voltage.

The pronounced ferroelectric-like hysteresis loops of the bending-voltage curves and current maxima on the double hysteresis current-voltage curves is typical for the electron-open electrodes. The analytical expression proving that the bending can be detected by interferometric methods is derived. The current maximum positions on the current-voltage loop are very close to the zero bending voltage on the bending-voltage hysteresis. Pronounced double hysteresis of current-voltage loop indicates the possibility of resistive switching between high and low resistance states. The switching occurs due to the strong coupling between electronic and ionic charge transfer appeared under the changes of applied voltage. The sharp meta-stable maximum of the electron density forms near one electrode and moves to another one during the periodic change of applied *ac* voltage.

The results can explain the nonlinear nature and correlation of electrical and mechanical memory effects in thin films of MIECs and solid electrolytes.

**Acknowledgements**

A.N.M. and E.A.E. acknowledge the support via bilateral SFFR-NSF project (US National Science Foundation under NSF-DMR-1210588 and State Fund of Fundamental of Fundamental Research of Ukraine, grant UU48/002). S.V.K. and A.T. acknowledges Office of Basic Energy Sciences, U.S. Department of Energy.

## Supplementary Materials
### Appendix S1. Bending of thin plate

Quasi-static solution for the problem is available for mechanically free film and could be expanded to the case of film clamped to a substrate. Let us consider the set of equations of state, relating stress components, $\sigma_{ij}$, with strain components, $u_{ij}$, for "active" film, containing inhomogeneous distribution of movable species with concentration $\delta N$

$$\begin{cases} u_{11} = \dfrac{\sigma_{11} - \nu_f(\sigma_{22} + \sigma_{33})}{Y_f} + W_{11}^d \delta N, \\ u_{22} = \dfrac{\sigma_{22} - \nu_f(\sigma_{11} + \sigma_{33})}{Y_f} + W_{22}^d \delta N, \\ u_{33} = \dfrac{\sigma_{33} - \nu_f(\sigma_{22} + \sigma_{11})}{Y_f} + W_{33}^d \delta N, \\ u_{23} = \dfrac{(1+\nu_f)\sigma_{23}}{Y_f}, \quad u_{13} = \dfrac{(1+\nu_f)\sigma_{13}}{Y_f}, \quad u_{12} = \dfrac{(1+\nu_f)\sigma_{12}}{Y_f}. \end{cases} \quad (S1.1)$$

Here $Y$ and $\nu$ are Young modulus and Poisson coefficient respectively, while $W_{ij}^d$ are components of Vegard strain tensor. These equations must be supplemented by the equilibrium conditions of bulk and surface forces, namely, $\partial\sigma_{ij}/\partial x_j=0$ in the bulk and $\sigma_{ij}n_j|_S=0$ at the free surface of the system [56]. Elastic stress should satisfy the equation of state (S1.1) and the mechanical equilibrium equation in the bulk

$$\frac{\partial \sigma_{ij}}{\partial x_i} = 0. \quad (S1.2)$$

And the boundary conditions at free surfaces

$$\sigma_{ij}n_j|_S = 0 \quad (S1.3)$$

Considering one-dimensional distributions it is more convenient to find strain and stress fields directly, without introducing displacement components. In this case equations (S1.3) and (S1.4) reduces to $\partial \sigma_{i3}/\partial x_3 = 0$ and $\sigma_{i3}|_S=0$ (with $i=1, 2, 3$). Thus, components $\sigma_{i3}$ are zero throughout the sample. In this case Eq.(S1.1) could rewritten as

$$\begin{cases} \sigma_{11} = \dfrac{Y_f}{1-\nu_f^2}(u_{11} - W_{11}^d \delta N) + \dfrac{Y_f \nu_f}{1-\nu_f^2}(u_{22} - W_{22}^d \delta N), \\ \sigma_{22} = \dfrac{Y_f}{1-\nu_f^2}(u_{22} - W_{22}^d \delta N) + \dfrac{Y_f \nu_f}{1-\nu_f^2}(u_{11} - W_{11}^d \delta N), \\ u_{33} = W_{33}^d \delta N - \dfrac{\nu_f(u_{22} - W_{22}^d \delta N + u_{11} - W_{11}^d \delta N)}{1-\nu_f}. \end{cases} \quad (S1.4)$$

Next one could recall the conditions of elastic compatibility $\text{inc}(i,j,\hat{u}) = e_{ikl}e_{jmn}\partial u_{ln}/x_{km} = 0$. For the considered case they could be reduced to $\partial^2 u_{11}/\partial x_3^2 = 0$, $\partial^2 u_{22}/\partial x_3^2 = 0$ (while the distribution of $u_{33}$ can be arbitrary along $x_3$). Obvious solution is

$$u_{11} = u_{11}^{(0)} + \frac{x_3}{R_1} \tag{S1.5}$$

$$u_{22} = u_{22}^{(0)} + \frac{x_3}{R_2} \tag{S1.6}$$

Constants $u_{11}^{(0)}$, $u_{22}^{(0)}$ and radii of curvature $R_1$ and $R_2$ should be found from the boundary conditions for corresponding stress component on free surfaces and continuity of displacements on the interface between two parts. Also one should consider the Saint-Venant approximation for the boundary conditions on the edges of the system

$$\int_0^h \sigma_{11}(x_3)dx_3 = 0, \quad \int_0^h \sigma_{11}(x_3)x_3 dx_3 = 0$$
$$\int_0^h \sigma_{22}(x_3)dx_3 = 0, \quad \int_0^h \sigma_{22}(x_3)x_3 dx_3 = 0 \tag{S1.7}$$

which means zero total force and total moment acting on the system in its plane. For the case when the "active" film with mobile ions is much thicker/stiffer than other "passive" layers (electrodes, substrate etc.), the solution for the strain distribution could be obtained from Eqs.(S1.5), (S1.6) and (S1.7), namely:

$$u_{11} = W_{11}^d \frac{1}{h}\int_0^h \delta N(z)dz + \frac{x_3}{R_1}, \quad u_{22} = W_{22}^d \frac{1}{h}\int_0^h \delta N(z)dz + \frac{x_3}{R_2} \tag{S1.8a}$$

$$u_{33} = W_{33}^d \delta N - \frac{\nu(u_{22} - W_{22}^d \delta N + u_{11} - W_{11}^d \delta N)}{1-\nu} =$$
$$= W_{33}^d \delta N - \frac{\nu(W_{11}^d + W_{22}^d)}{1-\nu}\left(\frac{1}{h}\int_0^h \delta N(z)dz - \delta N\right) - \frac{\nu}{1-\nu}\left(\frac{1}{R_1} + \frac{1}{R_2}\right)x_3 \tag{S1.8b}$$

Where the following designations relations are obtained

$$\frac{1}{R_1} = W_{11}^d \frac{12}{h^3}\int_0^h \left(z - \frac{h}{2}\right)\delta N(z)dz, \quad \frac{1}{R_2} = W_{22}^d \frac{12}{h^3}\int_0^h \left(z - \frac{h}{2}\right)\delta N(z)dz. \tag{S1.9}$$

Here $R_1$ and $R_2$ have the meaning of curvature radiuses of the plate surface.

From the strain field one could derive the following displacement components

$$u_1(x_1,x_2,x_3) = \left(W_{11}^d \frac{1}{h}\int_0^h \delta N(z)dz + \frac{x_3}{R_1}\right)x_1, \quad u_2(x_1,x_2,x_3) = \left(W_{22}^d \frac{1}{h}\int_0^h \delta N(z)dz + \frac{x_3}{R_2}\right)x_2,$$
$$u_3(x_1,x_2,x_3) = \int_0^{x_3} u_{33}(\tilde{x}_3)d\tilde{x}_3 - \frac{1}{2}\left(\frac{x_1^2}{R_1} + \frac{x_2^2}{R_2}\right). \tag{S1.10}$$

## Appendix S2. Coupled equations in dimensionless variables

Using the evident expression for the elastic stress (11), one can rewrite the coupled equations (8)-(9) for chemical potentials $\tilde{\mu}_d$ and $\tilde{\mu}_e$ in dimensionless variables

$$\frac{t_d}{t_e}\frac{\partial(f(\tilde{\mu}_d - \tilde{E}_d))}{\partial \tilde{t}} - \frac{\partial}{\partial \tilde{z}}\left(f(\tilde{\mu}_d - \tilde{E}_d)\frac{\partial}{\partial \tilde{z}}(\tilde{\varphi} - \tilde{\mu}_d + \tilde{w}^2 f(\tilde{\mu}_d - \tilde{E}_d))\right) = 0, \quad \text{(S2.1a)}$$

$$\frac{\partial}{\partial \tilde{t}}\left(F_{1/2}(\tilde{\mu}_e - \tilde{E}_C)\right) - \frac{\partial}{\partial \tilde{z}}\left(F_{1/2}(\tilde{\mu}_e - \tilde{E}_C)\frac{\partial}{\partial \tilde{z}}(\tilde{\mu}_e - \tilde{\varphi})\right) = 0. \quad \text{(S2.1b)}$$

Poisson equation (1) for electric potential $\tilde{\varphi}$ in dimensionless variables acquires the form:

$$\frac{\partial^2 \tilde{\varphi}}{\partial \tilde{z}^2} = -\left(\frac{N_d^0}{\overline{N}_d^+}f(\tilde{\mu}_d - \tilde{E}_d) - \frac{N_C}{\overline{N}_d^+}F_{1/2}(\tilde{\mu}_e - \tilde{E}_C)\right). \quad \text{(S2.2)}$$

One could get dimensionless donors concentration and electron density as $\tilde{N} = (N_d^0/\overline{N}_d^+)f(\tilde{\mu}_d - \tilde{E}_d)$ and $\tilde{n} = (N_C/\overline{N}_d^+)F_{1/2}(\tilde{\mu}_e - \tilde{E}_C)$. Approximation for "direct and "inverse" Fermi integrals are $F_{1/2}(\varepsilon) \approx \left(\exp(-\varepsilon) + (3\sqrt{\pi}/4)(4 + \varepsilon^2)^{-3/4}\right)^{-1}$ and $F_{1/2}^{-1}(\tilde{n}) \approx (3\sqrt{\pi}\tilde{n}/4)^{2/3} + \ln(\tilde{n}/(1+\tilde{n}))$ correspondingly.

Boundary conditions for currents in dimensionless variables are regarded donor-blocking and electron-conducting at least partially:

$$\tilde{J}_d\big|_{\tilde{z}=0} = 0; \quad \tilde{J}_d\big|_{\tilde{z}=\tilde{h}} = 0, \quad \text{(S2.3)}$$

$$\left(\tilde{J}_e - \tilde{\xi}_0(\tilde{n} - \tilde{n}_b)\right)\big|_{\tilde{z}=0} = 0 \quad \left(\tilde{J}_e + \tilde{\xi}_h(\tilde{n} - \tilde{n}_b)\right)\big|_{\tilde{z}=\tilde{h}} = 0 \quad \text{(S2.4)}$$

The currents are $\tilde{J}_e = \tilde{n}(\partial(\tilde{\mu}_e - e\tilde{\varphi})/\partial \tilde{x}_3)$ and $\tilde{J}_d = -(\eta_d/\eta_e)\tilde{N}(\partial(eZ_d\tilde{\varphi} - \tilde{\mu}_d - \tilde{W}\tilde{\sigma})/\partial \tilde{x}_3)$, where $\tilde{W}\tilde{\sigma} = -\tilde{w}^2(\tilde{N} - 1)$ and $\tilde{w}^2 = \dfrac{2W^2\overline{N}_d^+}{(s_{11} + s_{12})k_B T}$. Boundary conditions for sinusoidal electric potential are

$$\tilde{\varphi}\big|_{\tilde{z}=0} = \tilde{V}\sin(\tilde{\omega}\tilde{t}), \quad \tilde{\varphi}\big|_{\tilde{z}=\tilde{h}} = 0. \quad \text{(S2.4)}$$

Dimensionless variables and parameters involved in Eqs.(S2.1)-( S2.4) are listed in the **Table S2.1.**

**Table S2.1. Dimensionless variables and parameters**

| Quantity | Definition/designation |
| --- | --- |
| Dimensionless coordinate and thickness | $\tilde{z} = x_3/L_D$, $\tilde{h} = h/L_D$ |
| Debye screening length | $L_D = \sqrt{\varepsilon_0 \varepsilon_{33}^b k_B T/(e^2 \overline{N}_d^+)}$ |
| Dimensionless time | $\tilde{t} = t/t_e$ |
| Dimensionless frequency of applied voltage | $f = t_e \omega/2\pi$ |
| Characteristic electronic and donor | $t_e = L_D^2/(e\eta_e k_B T)$, $t_d = L_D^2/(e\eta_d k_B T)$ |
| times | |
| Dimensionless donor concentration | $\widetilde{N} = N_d^+/\overline{N}_d^+ = (N_d^0/\overline{N}_d^+)f(\widetilde{\mu}_d - \widetilde{E}_d)$ |
| Dimensionless electron density | $\widetilde{n} = n/\overline{N}_d^+ = (N_C/\overline{N}_d^+)F_{1/2}(\widetilde{\mu}_e - \widetilde{E}_C)$ |
| Equilibrium concentration of ionized donors at zero potential and stress | $\overline{N}_d^+ = N_d^0(1 - f((E_d - E_F)/k_BT))$ |
| Effective density of states in the conductive band | $N_C = \left(\dfrac{m_n k_B T}{2\pi\hbar^2}\right)^{3/2}$ |
| Dimensionless electric potential | $\widetilde{\varphi} = e\varphi/k_BT$ |
| Applied voltage | $\widetilde{V} = eU/k_BT$ |
| Dimensionless chemical potentials | $\widetilde{\mu}_d = \mu_d/k_BT$, $\widetilde{\mu}_e = \mu_e/k_BT$ |
| Dimensionless donor level | $\widetilde{E}_d = E_d/k_BT$ |
| Dimensionless conduction band position | $\widetilde{E}_C = E_C/k_BT$ |
| Dimensionless Fermi energy | $\widetilde{E}_F = E_F/k_BT$ is determined self-consistently from the electroneutrality $N_C F_{1/2}(\widetilde{E}_F - \widetilde{E}_C) = N_d^0 f(\widetilde{E}_F - \widetilde{E}_d)$ |
| Dimensionless Vegard coefficient | $\widetilde{w}^2 = \dfrac{2W^2\overline{N}_d^+}{(s_{11} + s_{12})k_BT}$ |
| Dimensionless electron and donor currents | $\widetilde{J}_e = \dfrac{L_D J_e}{e\overline{N}_d^+ \eta_e k_BT}$, $\widetilde{J}_d = \dfrac{L_D J_d}{e\overline{N}_d^+ \eta_e k_BT}$, |
| Total electric current is the sum of electronic, donor and displacement components | $\widetilde{J} = \widetilde{J}_e + \widetilde{J}_d + \dfrac{\partial \widetilde{E}}{\partial \widetilde{t}}$ |
| Dimensionless rate constant and concentration | $\widetilde{\xi}_{d,e} = \dfrac{L_D v_{d,e}}{e\eta_e k_BT}$, $\widetilde{n}_b = \dfrac{n_b}{\overline{N}_d^+}$ |

Numerical simulations, performed in MathLab, demonstrated that the calculated physical variables, such as space charge density, total space charge, potential, electric field, current-voltage and bending-voltage curves are governed by several factors. Namely, quasi-static, slow or fast kinetic regimes are governed by applied voltage frequency $f$. Symmetric or asymmetric, in the sense of electron-blocking, partially or completely electron-open electrodes can be simulated by the rate constant $\xi_i$. Parameters ranges used in numerical simulations are listed in the **Table S2.2.**

**Table S2.2. Parameters used in numerical simulations**

| Parameter or ratio | Numerical values and comments |
|---|---|
| $\widetilde{h}$ | 20, 40 and 10 |
| $t_d/t_e \equiv \eta_e/\eta_d$ | 10, $10^2$, $10^3$ |
| $f$ | $10^{-1}$, $10^{-2}$ (very high and high frequencies denoted as $f_1$ and $f_2$) $10^{-3}$, $3\times10^{-4}$, $10^{-4}$ (intermediate frequencies denoted as $f_3 - f_5$) $10^{-5}$, $10^{-6}$ (low and very low frequencies denoted as $f_6$ and $f_7$) |
| $E_C$ (eV) | 0 (regarded as zero energy level, valence band energy is negative) |
| $E_d$ (eV) | −0.1 (donors are regarded enough shallow) |
| $E_F$ (eV) | Determined self-consistently from the electroneutrality condition |
| $m_n/m_0$ | 0.5 (lighter than free electron mass $m_0$) |
| $W$ (Å$^3$) | 10 (typical value within the range $(1 - 50)$Å$^3$) |
| $s_{11} + s_{12}$ (Pa$^{-1}$) | $3.44 \times 10^{-12}$ (typical value for solids) |
| $N_d^0$ (m$^{-3}$) | $10^{24}$ (typical value for solid electrolytes) |
| $T$ (K) | 298 (room temperature) |
| ε | 5 (background permittivity of MIEC) |
| $\tilde{n}_b$ | 1 (equal to the equilibrium Fermi level in the system in order to exclude the rectification effect in the static case) |
| $\tilde{\xi}_{0,h}$ | $10^3$ (for almost electron-open electrodes)<br>1 and $10^{-3}$ (for partially and weakly electron-open electrodes)<br>0 (for electron-blocking electrode) |